# Coherent Emission from Surface Josephson Plasmons in Striped Cuprates

D. Nicoletti[a,*], M. Buzzi[a], M. Fechner[a], P. E. Dolgirev[b], M. H. Michael[a,b], J. B. Curtis[b,c], E. Demler[d], G. D. Gu[e], and A. Cavalleri[a,f]

[a] *Max Planck Institute for the Structure and Dynamics of Matter, 22761 Hamburg, Germany*
[b] *Department of Physics, Harvard University, Cambridge, Massachusetts 02138, USA*
[c] *John A. Paulson School of Engineering and Applied Sciences, Harvard University, Cambridge, MA 02138, USA.*
[d] *Institute for Theoretical Physics, ETH Zurich, 8093 Zurich, Switzerland*
[e] *Condensed Matter Physics and Materials Science Department, Brookhaven National Laboratory, Upton, NY, USA*
[f] *Department of Physics, Clarendon Laboratory, University of Oxford, Oxford OX1 3PU, United Kingdom*
\* *e-mail: daniele.nicoletti@mpsd.mpg.de*

**The interplay between charge order and superconductivity remains one of the central themes of research in quantum materials. In the case of cuprates, the coupling between striped charge fluctuations and local electromagnetic fields is especially important, as it affects transport properties, coherence and dimensionality of superconducting correlations. Here, we study the emission of coherent terahertz radiation in single-layer cuprates of the $La_{2-x}Ba_xCuO_4$ family, for which this effect is expected to be forbidden by symmetry. We find that emission vanishes for compounds in which the stripes are quasi-static, but is activated when *c*-axis inversion symmetry is broken by incommensurate or fluctuating charge stripes, such as in $La_{1.905}Ba_{0.095}CuO_4$ and in $La_{1.845}Ba_{0.155}CuO_4$. In this case, terahertz radiation is emitted by surface Josephson plasmons, which are generally dark modes, but couple to free space electromagnetic radiation because of the stripe modulation.**

Nonlinear terahertz (THz) spectroscopy has recently emerged as a new tool to study the microscopic properties of quantum materials, being susceptible to the symmetry of low energy degrees of freedom and complementing already existing nonlinear optical probes (1). For example, THz third harmonic generation was shown to be a sensitive probe of superfluid stripes, which do not couple to light at linear order but participate in higher order responses (2,3). As such, the study of THz nonlinear optics in presence of frustrated couplings provides new opportunities to explore the symmetry of quantum materials.

Here, we focus on THz emission from high-$T_C$ cuprates, and demonstrate how this method is highly sensitive to the spatial arrangement of the superconducting state and its interaction with *charge-stripe order*.

The emission of THz radiation from materials illuminated with femtosecond optical pulses (4,5,6,7) is generally enabled by two classes of mechanisms. The first mechanism, active in transparent non-centrosymmetric materials such as ZnTe or LiNbO$_3$, is based on optical rectification, where the second order nonlinear optical susceptibility causes a time dependent electrical polarization (8). The second mechanism relies on the excitation of time dependent charge currents, and is well documented for biased high-mobility semiconductors (8). A number of additional reports of coherent THz radiation have been made for complex quantum materials, typically related to the perturbation of electronic and magnetic interactions. THz emission in colossal magnetoresistance manganites (7,9,10), magnetic and multiferroic compounds (11,12,13,14,15,16,17,18,19,20) are some of the best-known examples.

In the case of high-$T_C$ superconductors, coherent THz emission has been reported only for situations in which time dependent supercurrents, $\dot{j}_s(t)$, are set in (8). These situations range from near-single-cycle THz pulses in biased antennas fabricated from YBa$_2$Cu$_3$O$_{7-\delta}$ or Bi$_2$Sr$_2$CaCu$_2$O$_{8+\delta}$ films (5,21,22), to multi-cycle narrowband emissions governed by the Josephson effect in the case of applied out-of-plane magnetic fields (23). It has also been shown that the use of Josephson junction stacks in MESA-type resonant structures allows orders of magnitude increase in THz emission efficiency, also providing narrow bandwidths and tuneable frequency (24,25,26).

Here, we report anomalous THz emission in high-$T_C$ cuprates, observed for photoexcitation with femtosecond near infrared pulses, in absence of external magnetic fields and current biases. The effect is detected only when superconductivity coexists with charge-stripe order in the Cu-O planes (27,28,29,30), and when these stripes are either incommensurate with the lattice or fluctuating.

We studied cuprates belonging to the "214" family, with one Cu-O layer per unit cell. As a prototypical "homogeneous" cuprate, we considered optimally-doped La$_{2-x}$Sr$_x$CuO$_4$ (LSCO), with a critical temperature of 38 K (see phase diagram in Fig. 1*A*). Although in the LSCO family fluctuating striped charge and spin orders have been reported in the underdoped region of the phase diagram (31), there is no evidence for stripes at optimal 0.16 doping (32). This sample was compared to the response of La$_{2-x}$Ba$_x$CuO$_4$ (LBCO), for which superconductivity coexists with charge stripes (27). We focused on three LBCO compounds: La$_{1.885}$Ba$_{0.115}$CuO$_4$ (LBCO 11.5%, $T_C$ = 13 K), where the superconducting transition is highly depleted by a robust stripe phase below the charge ordering temperature $T_{CO}$ = 53 K, La$_{1.845}$Ba$_{0.155}$CuO$_4$ (LBCO 15.5%, $T_C$ = 30 K, $T_{CO}$ = 40 K), placed at the nominal optimal doping and characterized by weak, highly fluctuating stripes (27), and La$_{1.905}$Ba$_{0.095}$CuO$_4$ (LBCO 9.5%, $T_C = T_{CO}$ = 33 K), for which the stripes have an intermediate intensity and correlation length compared to the other two compounds (27), but in contrast to them are here highly incommensurate (33,34). The location of the three samples in the LBCO phase diagram is shown in Fig. 1*B*-1*D*.

We note that $La_{2-x}Ba_xCuO_4$ is the same cuprate in which signatures of optically-enhanced superconductivity have been measured (35,36,37,38), and attributed to the ultrafast perturbation of the stripe order (39,40). In addition, a number of nonlinear optical effects, such as THz parametric amplification (41) and third harmonic generation (2), related to the resonant driving of Josephson plasma waves, have also been measured.

The main result of our experiment is summarized in Fig. 1*E*-1*L*, where the measured THz emission traces are reported for the four investigated compounds for selected temperatures, at a constant pump fluence of 2.5 mJ/cm². The experimental geometry is shown in the insets of the lower panels. We used the output of an amplified Ti:Sa laser as pump pulses, with a duration of 100 fs and photon energy of 1.55 eV (800 nm wavelength). These were focused at normal incidence onto an *ac*-oriented sample surface on a ~500 μm spot. The emitted THz pulses were collimated with a parabolic mirror and refocused on a 1-mm-thick ZnTe crystal to perform electro-optic sampling directly yielding THz electric field traces in time domain.

In optimally-doped LSCO (Fig. 1*E*), the THz emission signal was measurable only in the superconducting state below $T_C$, and displayed a very small amplitude, just above the noise level. This effect consisted of a single-cycle trace, with a flat and featureless spectrum (Fig. 1*I*). A similar response was also found in LBCO 11.5% (Fig. 1*F* and Fig. 1*J*), where charge stripes are robust, quasi-static and quasi-commensurate. Here, a barely detectable emission signal was also found for $T > T_C$.

On the other hand, in LBCO 15.5% (weak, highly fluctuating but quasi-commensurate stripes (33,34), Fig. 1*G* and Fig. 1*K*) the THz emission in the superconducting state acquired an appreciable amplitude, with oscillations at a frequency of ~600 GHz (depending on temperature).

In the compound with incommensurate, relatively strong stripes, *i.e.* LBCO 9.5%, the THz emission amplitude was even higher than LBCO 15.5% and greater by a factor of ~5-10 compared to LSCO and LBCO 11.5%. Coherent multi-cycle oscillations were observed (Fig. 1*H*), corresponding to a narrow spectral peak (Fig. 1*L*). The frequency of these oscillations shifted to the red with increasing temperature, whilst also reducing in amplitude and disappearing at $T_C$.

The rest of the analysis in this paper is focused on LBCO 9.5%, which yielded the largest signal and highest coherence. Firstly, we verified that the emission was entirely polarized along the out-of-plane crystallographic axis, and could be induced only for a pump polarization aligned along the same direction (see *SI Appendix*).

Figure 2*A* displays the pump fluence dependence measured at a constant temperature of 7 K. These experimental traces were modelled using fits in time domain (solid lines), for which we report the single components in the *SI Appendix*. These include a "single-cycle" pulse at early times, which was absent at the lowest fluences and grew quadratically with irradiation, and a quasi-monochromatic, long-lived oscillation, which grew linearly up to about 1 mJ/cm$^2$ and tended to saturate for higher excitation fluence (see Fig. 2*B*). This linear trend of the main oscillation is compatible with the impulsive excitation of a coherent mode. In the fluence-dependent behavior of lifetime and oscillation frequency (Fig. 2*C* and Fig. 2*D*), we identify a linear excitation regime where these quantities are weakly dependent on fluence and seem to stabilize at constant values of ~4 ps and ~0.45 THz, respectively. In this weak excitation regime, the driven mode parameters are well determined.

In Fig. 3 we report the temperature dependence of this effect. We show a comparison between the oscillation frequency in the THz emission signal in LBCO 9.5% and the bulk Josephson plasma resonance measured at equilibrium with time-resolved THz spectroscopy in the same sample. In the inset of Fig. 3*A* we show the experimental geometry, in which we illuminated the sample with weak broadband THz pulses (generated in a 200-μm-thick GaP), polarized along the out-of-plane direction, that were then detected in another 200-μm-thick GaP crystal via electro-optic sampling after being reflected from the sample surface.

Figure 3*A* displays examples of reflectivity ratios at two temperatures below $T_C$, normalized by the same quantity measured in the normal state. These curves evidence a Josephson plasma resonance, the exact frequency of which was determined by fitting the experimental data with a Josephson plasma model (solid lines) (35,38). The key result of this analysis is displayed in Fig. 3*B*, in which we show a comparison of the temperature dependence of the Josephson plasma frequency at equilibrium (gray) with the frequency of the emitted oscillations for two pump fluences. Notably, the emitted mode frequency hardens with decreasing fluence and approaches the equilibrium plasma frequency measured at the corresponding base temperature.

In interpreting our results, we first note that in a centrosymmetric cuprate impulsive excitation of Josephson plasmons is forbidden by symmetry. Josephson plasma modes are in fact symmetry-odd (infrared-active), while impulsive photo-excitation couples only to totally symmetric modes (42). As discussed in a related theory work (43), a prerequisite for the excitation of these modes is that charge order breaks inversion symmetry. However, this does not happen for commensurate quasi-static stripes, as those expected for dopings $x \gtrsim 1/8$ (33,34), which exhibit a two-fold screw axis along the out-of-plane direction (see Fig. 4*A*) (44). A symmetry breaking is expected instead for incommensurate or highly fluctuating stripes, as in the case of LBCO 9.5% and LBCO 15.5%. Here, the charge order correlation length along the out-of-plane direction is of the order of one unit cell (27), resulting in a loss of the phase relation between stripes in next-nearest-neighbouring planes (see Fig. 4B).

Once inversion symmetry is broken, electromagnetic emission at a frequency $\omega \ll \omega_{pump}$ can result from rectification of the optical pulse. We associate the optically rectified drive for plasma oscillations with the excitation of a *shift current* (43,44) at the sample surface. This is expected to interact with modes at $\omega \simeq \omega_{JPR}$, of which one finds at least two: (1) a *bulk Josephson plasma polariton*, sustained by tunnelling supercurrents oriented in the $z$ (out-of-plane) direction and propagating along the $x$ (in-plane) direction and (2) a *surface Josephson plasmon*, also sustained by plasma oscillations in the $z$ direction, but localized at the surface of the material and propagating along $z$. The dispersion relations for these two modes are shown in Fig. 4C and Fig. 4D, respectively (45).

Radiation from bulk plasma polaritons (Fig. 4C), excited over a depth between ~200 nm (skin depth of the pump) and ~1 μm (46), would be expected to be broad in frequency and overdamped. This is because excitation by the near infrared pump covers a wide range of in-plane momenta, $q_x$, which in the first instance is limited only by the envelope bandwidth of the pump pulse (gray shading in Fig. 4C). The spectrum of Josephson plasmons would, in this case, also be independent of the details of the stripe order and of its correlation lengths, as is instead observed. Moreover, one would expect radiation at frequencies $\omega \gtrsim \omega_{JPR}$, in contrast to the experimental observation of a slightly redshifted emission with respect to the plasma frequency (see Fig. 3B).

Coherent narrowband emission by surface Josephson plasmons is instead more likely. Although the dispersion of these modes lies below the light cone and, hence, they are not expected to radiate into vacuum (see Fig. 4*D*), we argue here that Bragg scattering off the stripe order induces a backfolding, defined by the stripe wave vector, into a reduced Brillouin zone (dashed horizontal line in Fig. 4*D*). For this reason, these surface modes can radiate, much like a situation in which a fabricated corrugation would be used to achieve the coupling (47,48,49,50).

In the right panel of Fig. 4*D*, we report the emission spectrum calculated for a striped superconductor through the excitation of surface Josephson plasmons. As extensively discussed in our related theory work (43,44), in the presence of stripes, the pump pulse is expected to give origin to an *Umklapp* shift current, $J_U \cos(Q_{stripes} z)$, that is modulated in space by the stripe wave vector, $Q_{stripes}$. This naturally drives high-momenta surface plasmons, which can radiate out due to the aforementioned backfolding mechanism.

In summary, we have reported the observation of coherent THz emission just below the Josephson plasma frequency in cuprates for which the superconducting state coexists with stripes. We assigned this effect to the excitation of surface Josephson plasmons, which become Raman active due to the breaking of inversion symmetry induced by the stripes, and can radiate out thanks to the backfolding of their dispersion curve onto the light cone. Based on these findings, the characterization of coherent THz emission emerges as a sensitive method to unveil broken symmetry states which may not be detectable with other conventional techniques. Moreover, the absence of THz emission in LBCO 11.5%, where the stripes are more robust and quasi-static, may suggest a qualitative difference in the nature of charge and spin order between compounds that are in the vicinity of the commensurate 1/8 doping and those that are far from it.

**Acknowledgments**

The research leading to these results received funding from the European Research Council under the European Union's Seventh Framework Programme (FP7/2007-2013)/ERC Grant Agreement No. 319286 (QMAC). We acknowledge support from the Deutsche Forschungsgemeinschaft (DFG, German Research Foundation) via the excellence cluster 'CUI: Advanced Imaging of Matter' (EXC 2056 – project ID 390715994) and the priority program SFB925 (project ID 170620586). E. Demler acknowledges support from AFOSR-MURI: Photonic Quantum Matter award FA95501610323, DARPA DRINQS, the ARO grant "Control of Many-Body States Using Strong Coherent Light-Matter Coupling in Terahertz Cavities". J. B. Curtis is supported by the Quantum Science Center (QSC), a National Quantum Information Science Research Center of the U.S. Department of Energy (DOE), and by the Harvard Quantum Initiative. J.B.C. also acknowledges hospitality from the Max Planck Institute for Structure and Dynamics of Matter (MPSD, Hamburg), and ETH Zürich Institute for Theoretical Physics. Work at Brookhaven is supported by the Office of Basic Energy Sciences, Division of Materials Sciences and Engineering, U.S. Department of Energy under Contract No. DE-SC0012704.

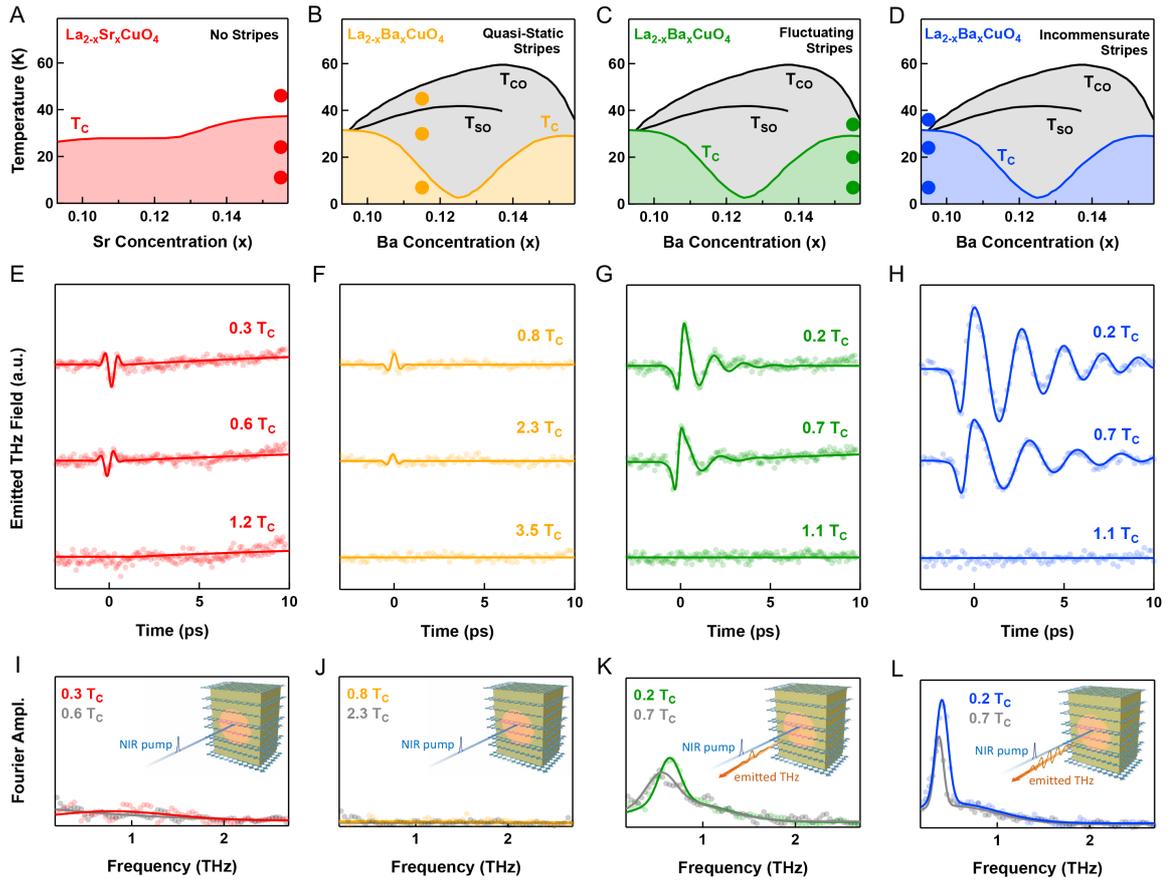

**Figure 1.** (*A-D*) Temperature-doping phase diagrams of the four compounds investigated in the present study. $T_{CO}$, $T_{SO}$, and $T_C$ stand for the charge ordering, the spin ordering, and the superconducting critical temperature, respectively. (*E-H*) Time-dependent THz emission traces taken for a pump fluence of 2.5 mJ/cm$^2$ at the temperatures indicated by full circles in (*A-D*). Solid lines represent multi-component fits to the data (see *SI Appendix*). The vertical scales in the three panels are mutually calibrated. (*I-L*) Fourier transforms (circles) of selected time-domain traces in (*E-H*). Solid lines are multi-Gaussian fits. Insets: Experimental geometry. Near-infrared (NIR) pump pulses are shone at normal incidence onto an *ac*-oriented sample surface, with polarization parallel to the *c* axis. As a result of photoexcitation, *c*-polarized THz radiation is emitted.

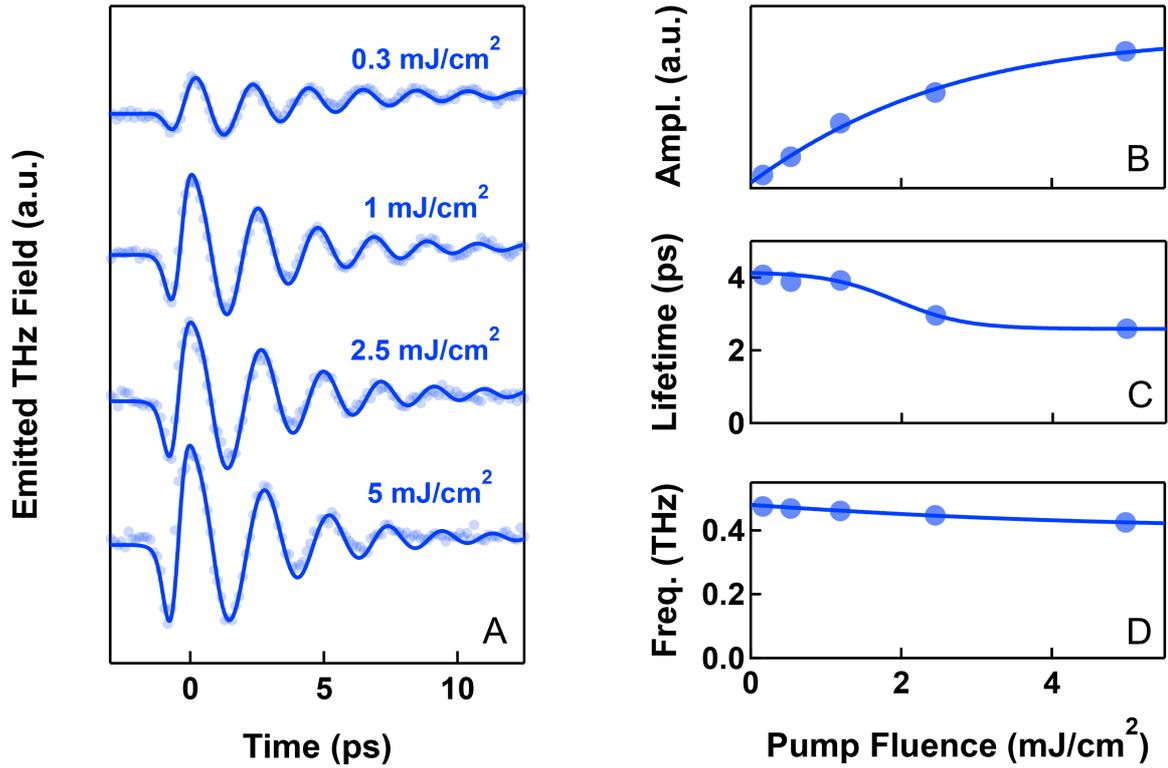

**Figure 2.** Pump fluence dependent THz emission in La$_{1.905}$Ba$_{0.095}$CuO$_4$ at $T$ = 7 K. (*A*) Experimental traces taken for different pump fluences (full circles). Solid lines are multi components fits to the data, which include a quasi-monochromatic, long-lived oscillation and a "single-cycle" component around time zero (see *SI Appendix*). (*B-D*) Fluence dependent parameters of the quasi-monochromatic oscillation extracted from the fits in (*A*).

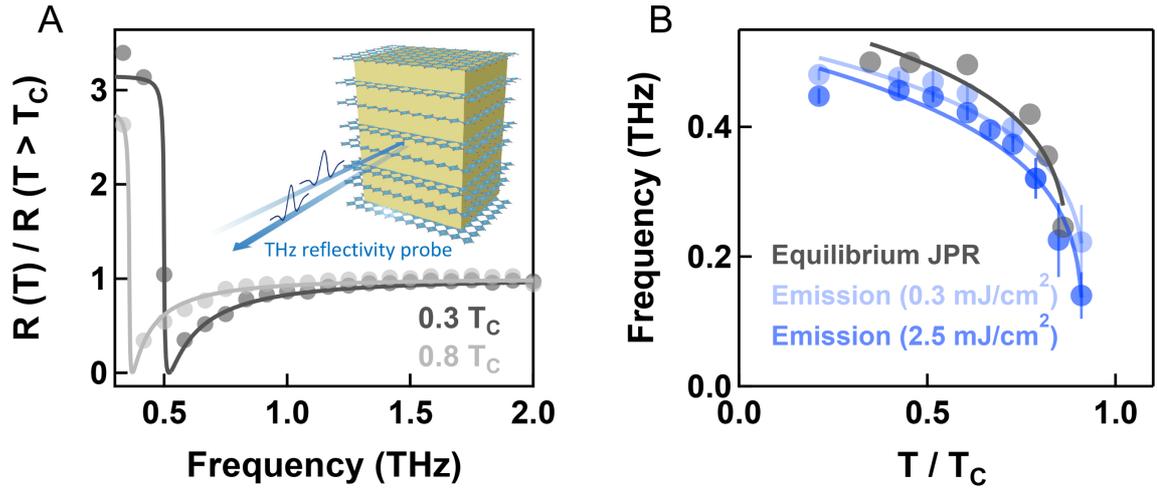

**Figure 3.** Comparison with the equilibrium Josephson plasma resonance in $La_{1.905}Ba_{0.095}CuO_4$. (*A*) Inset: Experimental geometry for the equilibrium THz time domain characterization. A weak broadband THz pulse was shone at normal incidence onto the sample surface with polarization along the *c* direction. The electric field profile of the same THz pulse was then detected after reflection. Main panel: Reflectivity taken at two different temperatures in the superconducting state, normalized by the same quantity measured at $T$ = 35 K > $T_C$ (full circles). The solid lines are fits to the data performed with a Josephson plasma model. (*B*) Temperature dependence of the equilibrium Josephson plasma frequency (gray circles), as determined from the fits in (*A*). The oscillation frequencies in the THz emission signal measured in the same sample are also reported for two different excitation fluences (see legend). Error bars indicate uncertainties extracted from fits such as those in Fig. 2 (see also *SI Appendix*). Solid lines are guides to the eye.

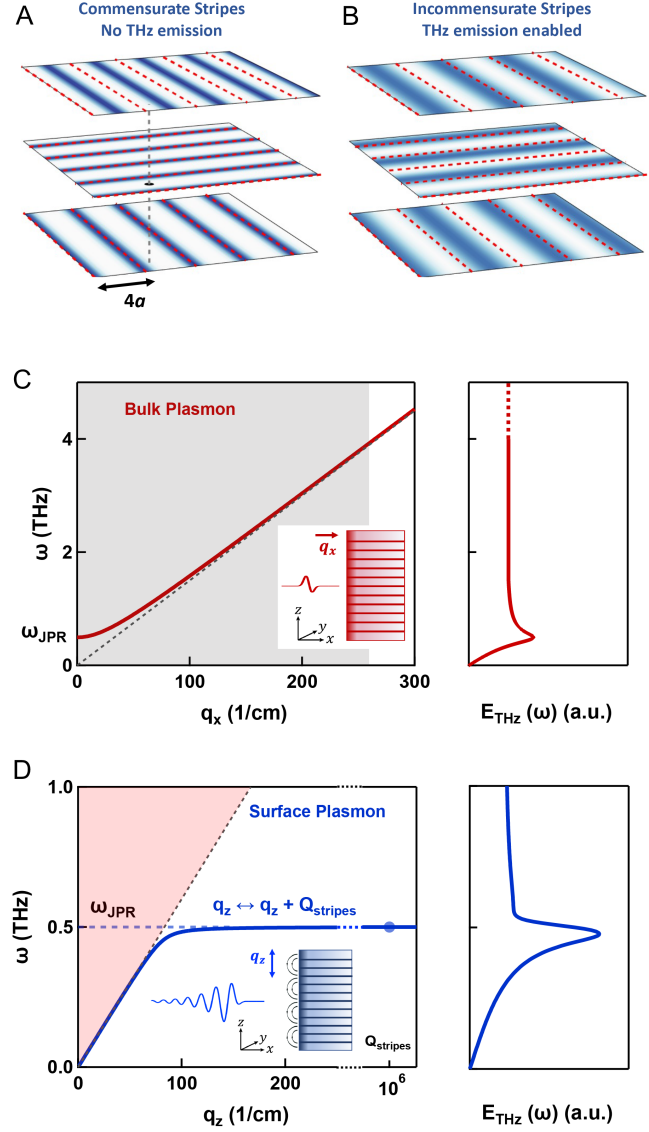

**Figure 4.** (*A*) Charge density pattern (gradual scale of blue) in three neighboring planes of a cuprate with commensurate stripes (red dashed lines are spaced by $4a$, where $a$ is the lattice constant). Stripes in next-nearest layers are off-phased by $\pi$ (30). Here, inversion symmetry is preserved (black dot and vertical dashed line are an inversion center and a screw axis, respectively). (*B*) Once commensurability is lost, stripes are fluctuating, or there is no phase relation between next-nearest layers, inversion symmetry can be broken and THz emission is enabled (43). (*C*) Left panel: In-plane dispersion of bulk Josephson plasma polaritons (red line). Emission from these modes (right panel) is expected to be very broad, as it encompasses a wide range of in-plane momenta, $q_x$ (gray shading) (43). (*D*) Left panel: Out-of-plane dispersion of surface Josephson plasmons (solid blue line). These modes are localized at the surface and propagate along $z$ (out-of-plane direction). As their dispersion lies below the light cone (red shading) they are not expected to radiate into vacuum. However, Bragg scattering off the stripe order induces a backfolding, defined by the stripe wave vector, $Q_{stripes}$, into a reduced Brillouin zone (dashed horizontal line). Hence, these surface modes are redirected into the light cone and can radiate out at frequencies just below $\omega_{JPR}(q=0)$. Right panel: Calculated emission spectrum from surface Josephson plasmons in a striped superconductor (43).

**REFERENCES (Main Text)**

# Coherent Emission from Surface Josephson Plasmons in Striped Cuprates

D. Nicoletti[a,*], M. Buzzi[a], M. Fechner[a], P. E. Dolgirev[b], M. H. Michael[a,b], J. B. Curtis[b,c], E. Demler[d], G. D. Gu[e], and A. Cavalleri[a,f]

[a] *Max Planck Institute for the Structure and Dynamics of Matter, 22761 Hamburg, Germany*
[b] *Department of Physics, Harvard University, Cambridge, Massachusetts 02138, USA*
[c] *John A. Paulson School of Engineering and Applied Sciences, Harvard University, Cambridge, MA 02138, USA.*
[d] *Institute for Theoretical Physics, ETH Zurich, 8093 Zurich, Switzerland*
[e] *Condensed Matter Physics and Materials Science Department, Brookhaven National Laboratory, Upton, NY, USA*
[f] *Department of Physics, Clarendon Laboratory, University of Oxford, Oxford OX1 3PU, United Kingdom*
*\* e-mail: daniele.nicoletti@mpsd.mpg.de*

# Supporting Information

**S1. Polarization dependence**

**S2. Pump pulse length dependence**

**S3. Fitting model**

**S4. Equilibrium Josephson plasma frequencies and detection bandwidth**

## S1. Polarization dependence

In Figure S1 we report a polarization dependent study. Data was taken on $La_{1.905}Ba_{0.095}CuO_4$ at $T = 7$ K after installing a tunable waveplate and an optical polarizer in the pump beam, as well as a THz polarizer in front of the detection crystal. As is evident from the traces, the multi-cycle THz emission signal at $\omega \simeq \omega_{JPR}$ appears to be entirely polarized along the out-of-plane crystallographic axis. Moreover, it is found only upon excitation with pump pulses polarized along the same *c*-axis direction (see Fig. S1d).

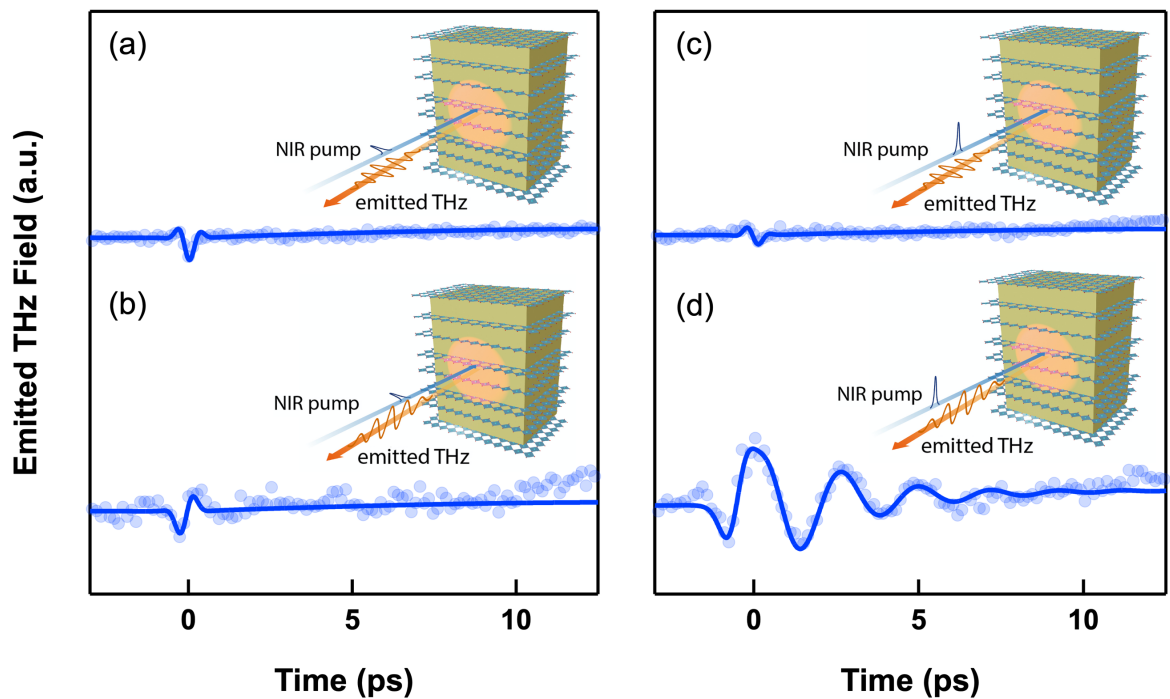

**Figure S1.** Pump and THz polarization dependence measured in $La_{1.905}Ba_{0.095}CuO_4$ at $T = 7$ K for a pump fluence of 2.5 mJ/cm². The following configurations are reported: **(a)** Pump and emitted THz both polarized in-plane, **(b)** Pump in-plane and emitted THz out-of-plane, **(c)** Pump out-of-plane and emitted THz in-plane, **(d)** Pump and probe both polarized out-of-plane. In each panel full circles are the experimental data while solid lines are multi-component fits.

This type of response, consistent with a dipolar distribution, allows us to rule out that the long-lived coherent oscillations originate from higher-order nonlinearities (electric quadrupole or magnetic dipole). These had been identified in Ref. (i) as responsible for the THz emission in unbiased $YBa_2Cu_3O_{6+x}$. Therein, however, the signal had very different characteristics from the one reported here, being present even in the normal state above $T_C$, and having a single-cycle shape and a $4\phi$ dependence with respect to the azimuthal angle $\phi$.

## S2. Pump pulse length dependence

As discussed in the main text, in a centrosymmetric cuprate impulsive excitation of Josephson plasmons is forbidden by symmetry.

One possibility is that photoexcitation leads to a direct coupling with another higher frequency fully symmetric mode that, in turn, can decay into Josephson plasmons. For example, an amplification of phase modes mediated by the amplitude mode[ii,iii,iv,v] has already been discussed in charge-density-wave materials[vi]. In Fig. S2 we report additional measurements in which we studied how THz emission in LBCO 9.5% evolved as the pump pulses were made longer, at constant fluence and temperature. The emitted oscillation amplitude reduced significantly as the pulse duration exceeded 2 ps [vii], pointing to a scenario in which modes at ~0.5 THz are excited directly, rather than indirectly by a high frequency symmetric mode.

The mechanism proposed in the main text, which involves the direct excitation of surface Josephson plasmons, is compatible with these results.

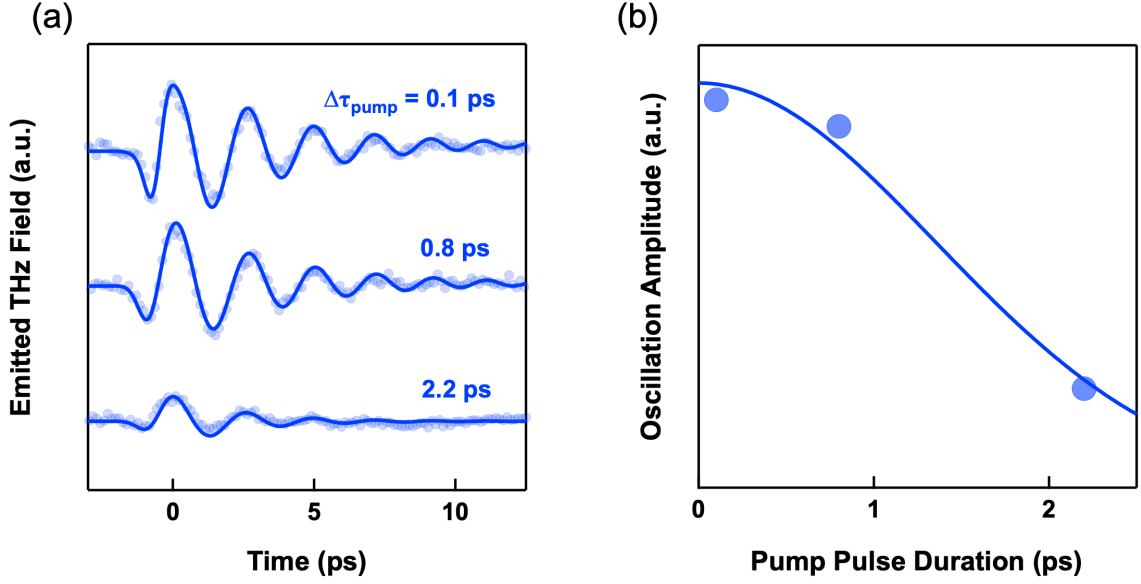

**Figure S2.** Pump pulse length dependence of the THz emission signal in $La_{1.905}Ba_{0.095}CuO_4$. **(a)** Experimental traces in time domain, taken at for different pump pulse durations, $\Delta\tau_{pump}$ (full circles). All data have been taken at $T = 7$ K, for a constant pump fluence of 2.5 mJ/cm². Solid lines are multi components fits. **(b)** Pulse length dependence of the THz oscillation amplitude (full circles), extracted from the fits in (a). The solid line is a guide to the eye.

## S3. Fitting model

All time-dependent experimental curves shown in Fig. 1 and Fig. 2 of the main text, as well as in Fig. S1 and Fig. S2 of the SI Appendix, were fitted using the formula:

$$E_{THz}(t) = A_0 e^{-(t/2\tau_0)^2} \cos(\omega_0 t + \varphi_0)$$
$$+ A_1[1 + \text{erf}(t/\tau_1)] e^{-\gamma_1 t} \cos[(\omega_1 + c_1 t)t + \varphi_1] + B(t)$$

Here, $A_0$, $\tau_0$, $\omega_0$, and $\varphi_0$ are the amplitude, Gaussian width, central frequency, and phase of the "single-cycle" component around time zero. $A_1$, $\tau_1$, $\gamma_1$, $\omega_1$, $c_1$ and $\varphi_1$ are instead amplitude, rise time, decay rate, initial frequency, linear chirp coefficient, and phase of the quasi-monochromatic, long-lived oscillation. $B(t)$ is a weak, slowly-varying polynomial background.

All fitting curves are displayed in Fig. 1, Fig. 2, Fig. S1, and Fig. S2 as solid lines. Note that for LSCO and LBCO 11.5% the experimental traces were reproduced using only a "single-cycle" component and weak a slowly-varying background (first and third terms in the equation above). In contrast, for LBCO 15.5% and LBCO 9.5% it was necessary to introduce the multi-cycle oscillation, which typically had much larger amplitude, $A_1$, than the other terms and only a weak (≲ 2%) positive linear chirp. Figure S3a illustrates such a fit for a particular set of data taken on LBCO 9.5%. The individual fit components are explicitly shown.

As reported in Fig. S3b, the "single-cycle" pulse was absent at low fluence and grew quadratically in amplitude with irradiation. On the other hand, the quasi-monochromatic, long-lived oscillation grew linearly up to about 1 mJ/cm² and saturated for higher excitation fluence. This linear trend of the main oscillation is compatible with the direct impulsive excitation of a coherent mode.

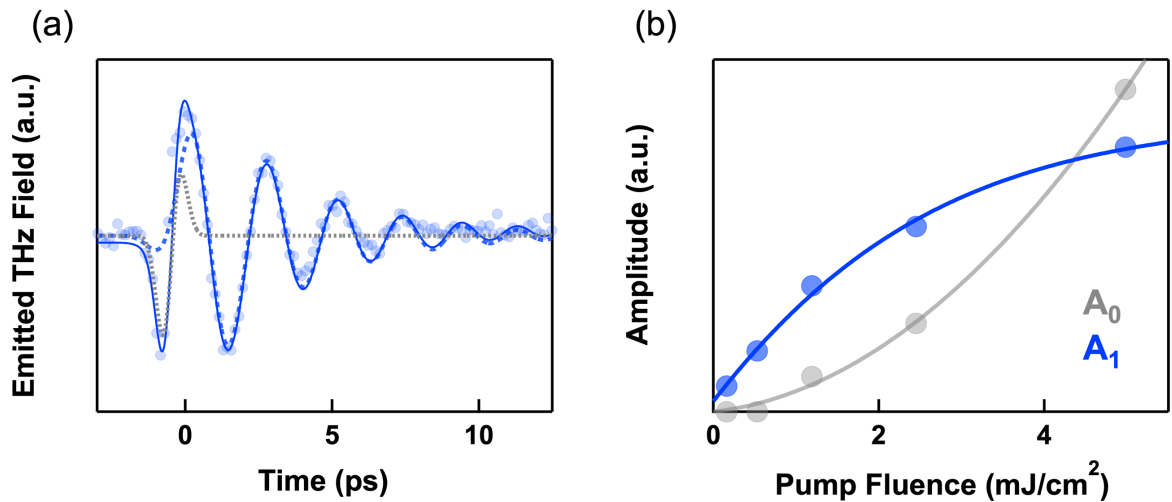

**Figure S3. (a)** THz emission signal measured in $La_{1.905}Ba_{0.095}CuO_4$ at $T$ = 7 K, for a pump fluence of 5 mJ/cm². Full circles are experimental data, while the blue solid line is the result of a multi-component fit. This included a "single-cycle" component around time zero (gray dots) and a quasi-monochromatic, long-lived oscillation (blue dashed line). **(b)** Fluence dependent amplitudes of both single-cycle (gray) and quasi-monochromatic (blue) components, extracted from fits as those shown in panel (a).

## S4. Equilibrium Josephson plasma frequencies and detection bandwidth

In Figure S4, we show the temperature dependence of the equilibrium Josephson plasma resonance for all four compounds investigated in our study. The spectra were measured by THz time-domain spectroscopy on the same crystals used in the emission experiment, and are in good agreement with previous reports in the literature[viii,ix,x].

Figure S5 shows instead the results of a measurement performed to estimate the bandwidth of our setup. We have installed a ZnTe crystal, identical to that used for detection of the emission signal, on the cryostat cold finger at the sample position. In Fig. S5a we report the THz electric field emitted by this ZnTe crystal, while Fig. S5b shows the corresponding spectrum. The frequency range shaded in red indicates a plausible detection bandwidth for our setup, estimated by setting the extremes to frequency values for which the spectral amplitude is reduced to 10% of the peak value. This extends over a range of ~0.1 to 2.75 THz, covering the Josephson plasma frequencies of all samples studied, which are indicated by arrows following the color coding of Fig. S4.

We stress here that although putative coherent emission from LBCO 11.5% could have occurred at frequencies lower than $\omega_{JPR} \simeq 0.2$ THz, we found a complete absence of long-lived oscillations in this material for any pump fluence down to values as low as 0.1 mJ/cm$^2$, for which, based on the LBCO 9.5% results of Fig. 2 and Fig. 3, emission should occur very close to $\omega_{JPR}$.

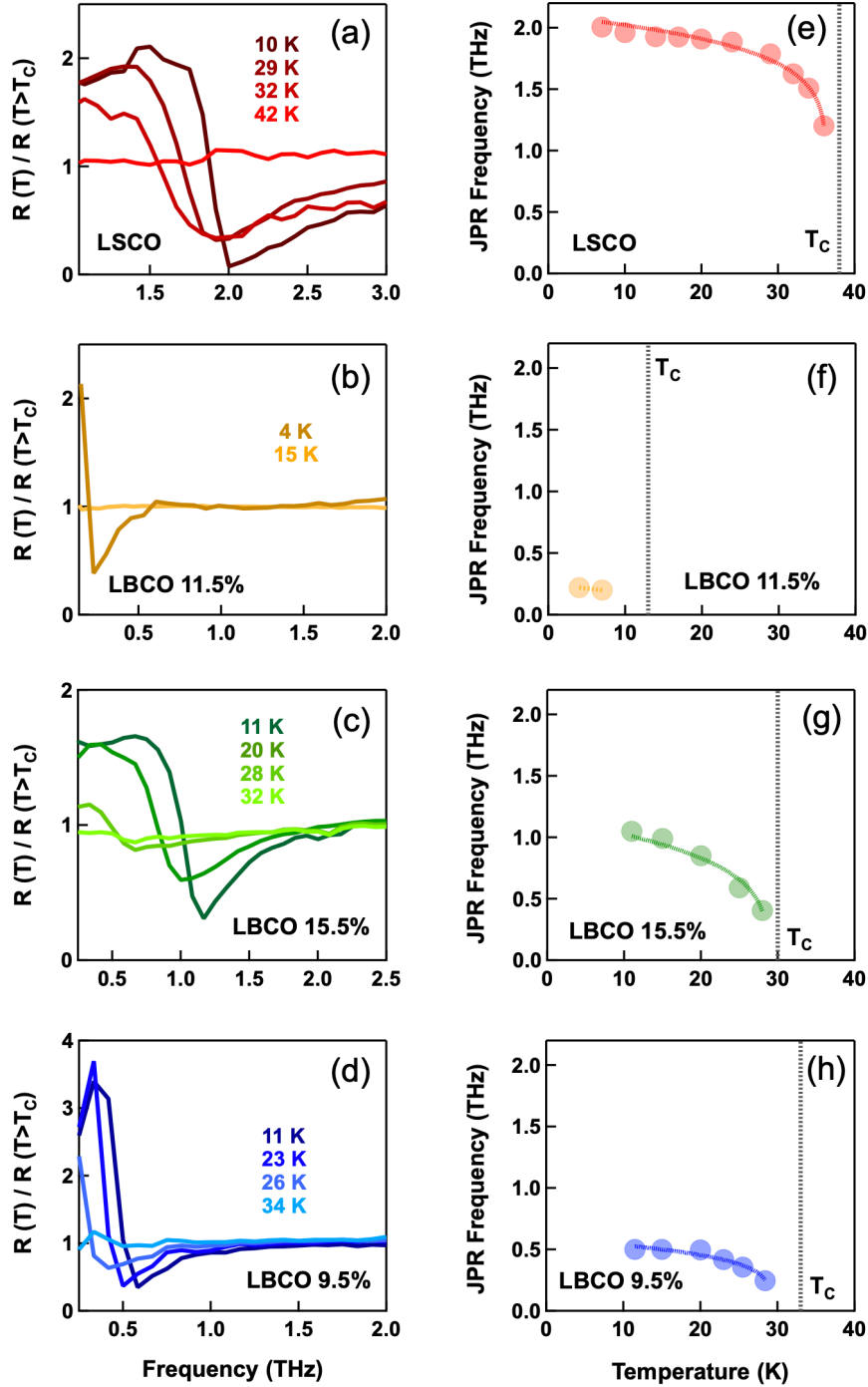

**Figure S4. (a-d)** Reflectivity of the four samples under study, normalized for the same quantity measured in the normal state above $T_C$. All spectra were measured by THz time-domain spectroscopy and are reported at selected temperatures below and across the superconducting transition. The reference temperatures for normalization were 50 K (LSCO), 25 K (LBCO 11.5%), and 40 K (LBCO 15.5% and LBCO 9.5%). The Josephson plasma resonance appears here as a sharp edge located at $\omega = \omega_{JPR}$. **(e-h)** Temperature dependence of $\omega_{JPR}$ extracted via fits on the reflectivity ratios in (a-d) with a Josephson plasma model (see Fig. 3a). The colored dashed lines are guides to the eye, while the vertical gray lines indicate the superconducting $T_C$ of each sample.

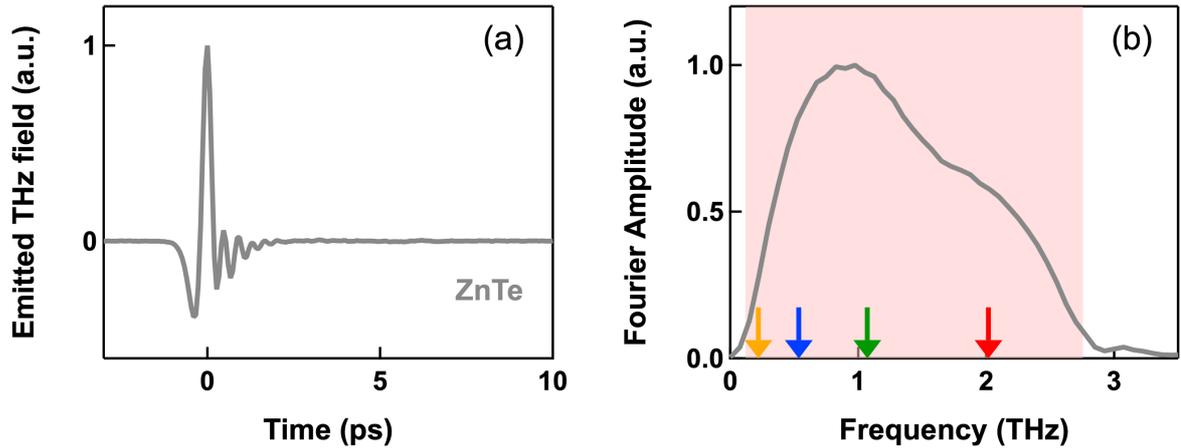

**Figure S5. (a)** Time-dependent electric field emitted by a 1-mm thick ZnTe crystal mounted on the cryostat cold finger at the sample position. The signal was measured by electro-optic sampling in an identical 1-mm thick ZnTe crystal, which was also used for all other measurements. **(b)** Fourier transform of the time trace in (a). The red shading indicates the spectral range between 0.1 and 2.75 THz, which is an estimate of the bandwidth of our experimental setup, having as extremes the frequencies for which the spectral amplitude is 10% of the peak value. Colored arrows indicate the Josephson plasma frequencies of the four samples in our study extracted at the lowest temperatures shown in Fig. S4. The color coding follows that of Fig. S4.

## REFERENCES (Supporting Information)